\documentclass[journal]{IEEEtran}
\usepackage{amsmath}    % need for subequations
\usepackage{graphics,epsfig,subfigure}    % need for figures
\usepackage{verbatim}   % useful for program listings
\usepackage{color}      % use if color is used in text

\usepackage{subfigure}  % use for side-by-side figures
\usepackage{hyperref}   % use for hypertext links, including those to exte$RN$al documents and URLs
\usepackage{amssymb}
\usepackage{epstopdf}
\usepackage{graphicx}
\graphicspath{{/Figures/}}
\usepackage{wasysym}
\usepackage{wrapfig}
\usepackage{sidecap}
\usepackage{float}
\usepackage{subfigure}
\usepackage{enumerate}
\usepackage{graphicx}
\graphicspath{{./figures/}}
\usepackage{algorithmic}
\usepackage{algorithm}
\usepackage{amsthm}
\usepackage{footnote}
\usepackage[official]{eurosym}
\usepackage[normalem]{ulem}
\makesavenoteenv{tabular}
\usepackage{multirow}

\newtheorem{theorem}{\textbf{Theorem}}

\newtheorem{definition}{Definition}
\newtheorem{problem}{Problem}

\begin{document}
\title{Vehicular Energy Network}

\author{Albert Y.S. Lam,
        Ka-Cheong Leung, and 
        Victor O.K. Li
\thanks{This work was supported in part by the Theme-Based Research Scheme of the Research Grants Council of Hong Kong under Grant T23-701/14-N. A.Y.S. Lam was also supported in part by the Seed Funding Programme for Basic Research of The University of Hong Kong under Grant 201601159009. \textit{(Corresponding author: Albert Y.S. Lam)}}		
\thanks{An earlier version of this paper was presented at the 6th IEEE International Conference on Smart Grid Communications, Miami, FL, November 2015 \cite{VEN_conf}.}
\thanks{The authors are with the Department of Electric and Electronic Engineering, The University of Hong Kong, Pokfulam Road, Hong Kong (e-mail: ayslam@eee.hku.hk; kcleung@eee.hku.hk; vli@eee.hku.hk).}
%\thanks{Manuscript received September 19, 2005; revised December 27, 2012.}
}

%\markboth{IEEE Transactions on Transportation Electrification}%,~Vol.~x, No.~x, xxx~20xx}%
%{ \MakeLowercase{Lam \textit{et al.}}: Vehicular Energy Network}

\maketitle

\begin{abstract}
%\boldmath

The smart grid spawns many innovative ideas, but many of them cannot be easily integrated into the existing power system due to power system constraints, such as the lack of capacity to transport renewable energy in remote areas to the urban centers. An energy delivery system can be built upon the traffic network and electric vehicles (EVs) utilized as energy carriers to transport energy over a large geographical region. A generalized architecture called the vehicular energy network (VEN) is constructed and a mathematically tractable framework is developed. Dynamic wireless (dis)charging allows electric energy, as an energy packet, to be added and subtracted from EV batteries seamlessly. With proper routing, energy can be transported from the sources to destinations through EVs along appropriate vehicular routes. This paper gives a preliminary study of VEN.  Models are developed to study its operational and economic feasibilities with real traffic data in the United Kingdom. Our study shows that a substantial amount of renewable energy can be transported from some remote wind farms to London under some reasonable settings and VEN is likely to be profitable in the near future. VEN can complement the power network and enhance its power delivery capability.

\end{abstract}
%\vspace{-2mm}
\begin{IEEEkeywords}
Dynamic charging, electric vehicle, energy routing, energy delivery.
\end{IEEEkeywords}

\IEEEpeerreviewmaketitle

%\vspace{-4mm}
%---------------------------------------------------------------------------
\section{Introduction}

The smart grid spawns many innovative ideas and applications, potentially revolutionizing the power system. Unfortunately, many of these new ideas cannot be easily integrated into the existing power system  due to power system constraints. One such constraint is the lack of capacity to transport renewable energy, typically in remote areas, to the urban centers. Any new smart grid design needs to be thoroughly tested and demonstrated to be fool-proof  before being integrated into the real power system. Some examples of complications are given below:

\subsubsection{Renewable Energy}
There is tremendous amount of renewable energy generation, but only a small portion is available and  utilized at the loads.  As of 2009, China had less than one-third of wind farms connected to the grid due to the difficulty of intermittent power dispatch and transmission network limitations \cite{ChinaRenewable}. In 2013, the ISO New England cut back power from wind and hydroelectric plants a few times because too much electricity was produced and transmission lines with robust carrying capacity  to connect the wind farms  located in remote areas were missing \cite{intermittentNature}. 
However, cutting  renewable output may lead to failure to meet the mandates of renewable energy generation in some countries \cite{futurePower}.

\subsubsection{Ancillary Services}
When discrepancies occur between supply and demand in the power system, ancillary services are required. Operating reserves can be provided in diverse locations of the grid and distributed energy resources (DERs), like the renewables and aggregations of EVs \cite{regulation_TSG}, may be  preferable over conventional generators in some situations. Further study is still required before  DERs may be brought into the power system effectively.

\subsubsection{New Energy Markets}
In the power system, the demand is connected with the supply through the power networks. The power infrastructure is generally managed or owned by multiple parties. However, the smart grid can introduce many variations and uncertainties to both supply and demand, e.g., from DERs, demand responses, vehicle-to-grid systems \cite{V2GMarket}. These new players have great potential to develop new energy markets with non-standard operational models, but they still rely  on the power network to transmit power. 
The grid operators in general oppose the integration of new operational models until they have undergone the reliability and security assurance. If the power network can be supplemented for energy transfer, the new energy businesses can thrive.

The common theme among the above scenarios is that the integration of new smart grid elements cannot guarantee that the operations of the current power system will not be affected.  If there exists a power delivery infrastructure independent of the conventional power network to connect the various (stable or unstable) types of power sources and loads, many of the brilliant ideas in smart grid can become a reality overnight. 

There is some related work toward that direction.
\cite{EVnetICC} proposed the EV energy network, which utilizes EVs for energy transmission and distribution.
In this design, EVs charge up when they stop at some energy routers and discharge the energy when stopping at other energy routers. In \cite{EVnetISGT}, a greedy algorithm for energy scheduling and allocation was developed. Shortest path routing \cite{EVnetMASS} and multiple path routing \cite{EVnetISGT2014} were also studied.
All these efforts are based on the model developed in \cite{EVnetICC} and a well-defined framework is required for more in-depth analysis.
Online Electric Vehicles (OLEVs) \cite{OLEV} are EVs which support dynamic charging, i.e., wireless charging in motion, and they have been commercialized in Korea. \cite{OLEV_wireless} designed a wireless power transfer (WPT) system to facilitate dynamic charging for OLEVs. \cite{OLEV_econ} analyzed the benefits of dynamic charging with an economic model and it showed that dynamic charging is beneficial to battery life.
\cite{MED1} and \cite{MED2} introduced mobile energy disseminators (MEDs). MED is a moving vehicle and it can wirelessly charge some other moving vehicles in the neighborhood.

In this paper, our preliminary work \cite{VEN_conf} is extended and we aim to develop a framework to model an energy delivery architecture for transporting energy from one place to another by means of EVs. Our framework generalizes the above mentioned infrastructure and designs and the system  modeled by this framework is called the vehicular energy network (VEN). This framework defines the necessary components  to facilitate more in-depth research. 
The rest of the paper is organized as follows: VEN with supporting arguments and characteristics are defined in Section \ref{sec:VEN}. Section \ref{sec:model} gives the system model and quantifies the transferable energy and loss. In Section \ref{sec:formulation}, the system is analyzed by studying  energy transfer maximization and energy loss minimization. 
An economic model is developed for VEN in Section \ref{sec:economic}.
Section \ref{sec:simulation}  studies the operational and economic feasibilities of VEN and the paper is concluded in Section \ref{sec:conclusion}.

%---------------------------------------------------------------------------
\section{System Design} \label{sec:VEN}

There are a number of existing technologies facilitating the establishment of VEN:
\subsubsection{Renewable Energy}
To confront global warming and climate change, many nations have established targets and mandates for renewable energy use \cite{targets}.  For example, California and Colorado in the U.S.  have mandated  renewable targets of 33\% and 30\% by 2020, respectively. The European Union 2030 target is at least 30\% of energy coming from renewable sources \cite{targets}. China sets the 15\% renewable target with 500 GW renewable electricity by 2020  \cite{targets}. Moreover, renewable power capacity is massive \cite{REN21}. Therefore, an appropriate approach to manage the abundant renewable energy generation can help meet various nations' energy mandates.

\subsubsection{Electric Vehicles}
EVs refer to a family of vehicles with batteries equipped to store energy for operations.
An EV can be considered as a \textit{movable energy storage} for various applications, e.g.,  facilitating frequency regulation \cite{regulation_TSG} and demand response \cite{TTE2}. Boosting the number of EVs is also included in the green policies of many countries. For example, the U.S. sets the goal of having one million EVs on the road by 2015 \cite{USEV} and China targets to have a similar goal \cite{ChinaEV}. Many automotive companies have already included EVs in their major production lines.
As the related equipment (e.g., batteries) is improved and the facilitates (e.g., charging stations \cite{EVCPP}) become available,  EVs will be prevalent in the near future.

\subsubsection{Vehicular Ad-hoc Network (VANET)}
VANET is a mature technology allowing vehicles, as mobile nodes, to communicate with each other or some fixed infrastructures \cite{VANET}.
Newly designed vehicles, especially EVs, are equipped with many sensors and this allows VANET to support a variety of functions, such as surveillance and route planning. This nurtures new applications where  vehicles submit (parts of) their intending traveling paths for planning purposes.

\subsubsection{Wireless Power Transfer}
Most near-field electromagnetic induction-based WPT techniques can be categorized into magnetic induction and electrostatic induction \cite{r2_4}: The former is suitable for applications with various power levels and gap separations while the latter is more restricted to those with small gap distances.
\cite{r2_2} reviewed various WPT technologies for EV wireless charging.
An inductively coupled multi-phase resonnant converter was designed for wireless EV charging applications \cite{TTE4}.
\cite{TTE1} focused on heavy duty vehicles and discussed the required elements of WPT for high power charging.  Furthermore, batteries of EVs can also be charged wirelessly on the move, i.e., dynamic charging. Dynamic charging facilitates frequent charging so that the batteries can be made smaller and the overhead costs due to batteries can be lowered. Shallow and frequent charging can enhance battery life \cite{EconDCEV}. 
A double-spiral repeater was proposed to design effective dynamic WPT systems \cite{TTE3}. \cite{cuttingCord} revealed that inductive power transfer is promising to dynamic charging. In \cite{coilDesign}, an efficient dynamic wireless charging system for EVs was designed based on magnetic coupled resonant power transmission.
\cite{r2_1 investigated power smoothing of dynamic wireless EV charging and addressed the various issues due to power pulsations. Some WPT designs do not require batteries. 
\cite{r2_3} discussed roadway-powered EVs which can obtain power from power supply rails under the road without storing the energy in batteries. A 1-MW inductive power transfer system was designed for high-speed trains to support dynamic charging without using batteries \cite{r2_5}.}
Some companies, like Qualcomm \cite{qualcomm}, ABB, and Microsoft \cite{ABB_MS}, are working toward improving the wireless charging technologies for EVs. A Stanford team proved that power up to 10 kW could be transferred effectively with a moving car \cite{stanford}. OLEVs have been deployed in Gumi City of South Korea \cite{EconDCEV}. England demonstrated a real dynamic WPT system on some motorways \cite{Englandmotorway}. In \cite{DCSim}, a simulation model was developed to analyze the daily amount of energy supported by a dynamic charging system. These evidences show that wireless charging and discharging can take place without interfering the movements of EVs.

The aforementioned technologies and developments make VEN a possible energy delivery infrastructure. 
EVs can be utilized  as \textit{energy carriers}. Consider an EV moving on a particular route connecting Locations A and B. If the EV is wirelessly charged at A and discharged at B, energy will be brought from A to B via the EV. Again, suppose that after a while, another EV passes through Location B and moves toward Location C. Then that EV can be charged and discharged at B and C, respectively, and this allows us to bring the energy to Location C. In this way,  if there are vehicular routes intersecting one another and some pass through the energy sources and destinations, the required energy can be transmitted from the sources to the destinations through VEN. VEN is formally defined as follows:

\begin{definition}[Vehicular Energy Network]
VEN is a vehicular network aiming to distribute energy in a region by means of EVs without tight time limitation. 
It is built upon a road network where EVs run on certain routes and there are dynamic (dis)charging facilities, with limited storage for  energy between each charge and discharge, installed along the roadway or at some road junctions. Nodes with energy to be delivered and received are denoted as the energy sources and destinations, respectively. During charging, a certain portion of energy is transmitted to an EV from a charging facility, and similarly, a portion of energy is transmitted from an EV  to a discharging facility during discharging. An EV carries a \textit{``packet'' of energy} along its route between its charge and discharge. By properly charging and discharging certain EVs at selected locations along the roadway, energy can be brought from the energy sources to the energy destinations.
\end{definition}

We do not need to strictly enforce energy supply and demand balance at all times in VEN. The design of VEN is not tied to particular types of energy generation and load. It is relatively easier to construct VEN than the traditional grid in some places, e.g., remote areas. It has an ability to store up energy, as an energy storage, but in a global sense. Depending on the applications, we have different ways of manipulating the energy stored in the network. Moving energy around is still a fundamental operation of VEN. To the best of our knowledge, we cannot find a counterpart which is directly comparable to VEN, in terms of characteristics and functionalities.

\begin{figure}[!t]
\centering
\includegraphics[width=3.2in]{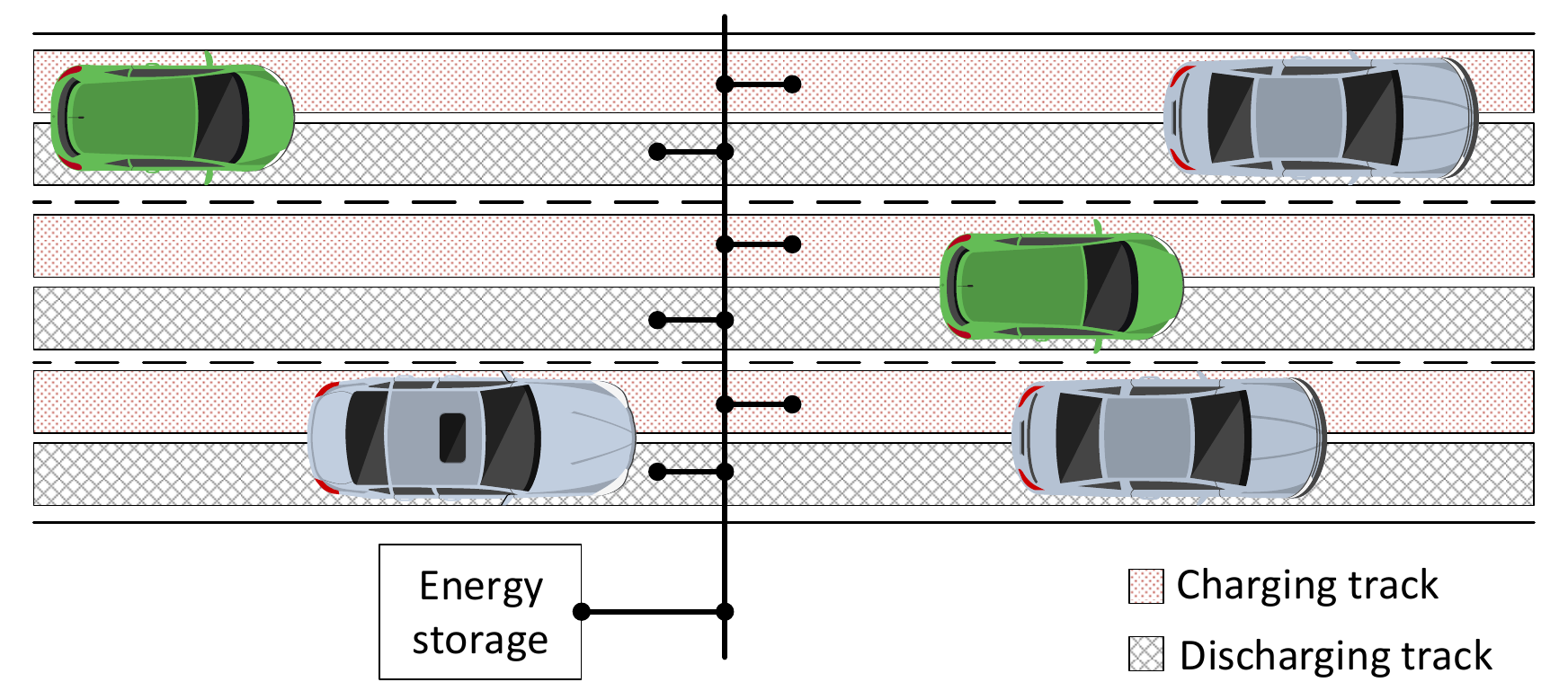}\vspace{-0.3cm}
\caption{Schematic of charging and discharging facilities with energy storage at a segment of road.}
\label{fig:track}
\vspace{-0.3cm}
\end{figure}
Fig. \ref{fig:track} gives an illustrative example of a segment of road, where charging and discharging tracks are buried underground in each designated lane. The charging and discharging coils can be installed at different locations to accommodate the configuration of roadway. These tracks are connected via a common bus with an energy storage attached. A (dis)charging track is responsible for (dis)charging the appropriate EVs running over it dynamically. The wireless energy transfer equipment can be designed similar to the one given in \cite{OLEV_wireless}. At any moment, it is possible to have multiple EVs undergoing charging and discharging simultaneously. Energy will be transferred from discharging EVs to charging EVs directly. The deficit will be compensated by the storage while the excess will be stored.

Note that we do not need to actively control the EVs participating in VEN. The EVs are not required to stop or slow down for (dis)charging at particular locations as energy can be transmitted  while they are moving. The EVs are not needed to follow any dedicated routes. Instead, those EVs are selected with favorable routes to carry energy so as to perform energy delivery.

\setcounter{subsubsection}{0}
VEN possesses the following characteristics which make it suitable as a supplement to the power grid to deliver power:
\subsubsection{Controllable energy transmission rate}
Energy is transmitted in the form of packets and this allows VEN to be managed like a packet-switched data network. Given the intended route of an EV,\footnote{For public transport, the exact schedules and routes can be acquired. For private vehicles, we expect that the participants are willing to disclose parts of their intended routes to the system. At the least, as soon as a vehicle is on a particular road segment, it will stay until the end of that segment for sure before it can exit.} It is known where the energy it carries can be delivered. By controlling how many EVs are utilized to carry energy and when proper charging and discharging events take place with the EVs, the energy flow can be controlled and the energy transmission rate can be specified on each road segment. 
\subsubsection{High flexibility}
A dedicated infrastructure does not need to be built in order to realize VEN. VEN relies on the existing road network, which covers  almost all locations involved in human activities. A road network can be transformed into a VEN when a certain number of (dis)charging facilities  have been installed. Unlike the conventional power network where the power sources and loads are generally pre-specified, the locations of  energy sources and destinations on VEN can be modified from time to time.
\subsubsection{Low overhead}
Equipment does not need to be specifically designed for VEN; even without VEN, WPT facilities will likely be found in the road network to serve the growing EV population in the future.

Existing technologies render VEN feasible. There are some non-technological factors which may be important when VEN is brought into reality. Due to space limitations, more details can be found in \cite{VEN_conf}.

\section{System Model} \label{sec:model}
\subsection{Road Network}
A road network is modeled with a directed graph $\mathcal{G}(\mathcal{N},\mathcal{A})$, where $\mathcal{N}$ and $\mathcal{A}$ are the set of energy points and the set of roads connecting the energy points, respectively. $\mathcal{N}$ includes all possible energy sources, destinations, and routing points set up at those locations with (dis)charging facilities installed. Depending on the application, $\mathcal{G}(\mathcal{N},\mathcal{A})$ may cover a district, a city, or even a whole country. Let $head(a_i)$ and $tail(a_i)$ be the head and tail of arc $a_i\in \mathcal{A}$. Each arc $a_i$ incurs a delay $d(a_i)$ time units for transferring energy from $tail(a_i)$ to $head(a_i)$; when energy is carried by a vehicle, it takes $d(a_i)$ units to traverse $a_i$. 

\subsection{Vehicular Traffic}
Assume that the traffic flows are static. Let $\mathcal{R}$ be the set of all possible vehicular routes in $\mathcal{G}$, each of which is loop-free. Each $r_i\in\mathcal{R}$ is a sequence of connected arcs, i.e., $r_i = \langle a_1',\ldots,a_{|r_i|}' \rangle$ with traffic flow $f_i$, which is the number of vehicles traveling on $r_i$ per unit time. The $n$-th arc of $r_i$ is denoted by $r_i(n)$. Let $r_i(n,m), n<m,$ be the sub-route of $r_i$ starting from the $n$-th arc and ending with the $m$-th arc. 

\subsection{Energy Path}
\begin{definition}[Energy Path]
An energy path $p(s,t)$ is the path along which energy is transmitted from the Energy Source Node $s$ to  the Destination Node $t$. Each path is composed of segments of the vehicular routes.
\end{definition}
For each pair of $(s,t)$, the set of all possible paths $\mathcal{P}(s,t)$ can be constructed. Each path $p_j(s,t)\in \mathcal{P}(s,t)$ can be represented by $p_j(s,t)=\langle r_1^j(n_1,m_1),\ldots, r_i^j(n_i,m_i),\ldots, r_{|p_j|}^j(n_{|p_j|},m_{|p_j|}) \rangle$, where $r_i^j(n_i,m_i)$ refers to the $i$-th segment of $p_j$, and it is the sub-route of $r_i^j$ with the starting arc $n_i$ and the ending arc $m_i$. $|p_j|$ is the number of sub-routes used to construct $p_j$. $p_j(s,t)$ also needs to satisfy
(i) $tail(r_1^j(n_1))=s$,
(ii) $head(r_{|p_j|}^j(m_{|p_j|}))=t$, and
(iii) $head(r_{i}^j(m_{i}))=tail(r_{i+1}^j(n_{i+1})),$ for $i=1,2,\ldots,|p_j|-1$.
\begin{figure}[!t]
\centering
\includegraphics[width=2.2in]{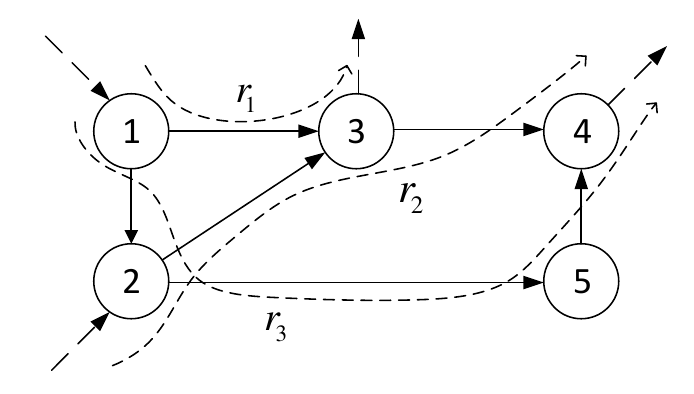}\vspace{-0.5cm}
\caption{Energy paths.}
\label{fig:energypaths}
\vspace{-0.3cm}
\end{figure}
Consider the example given in Fig. \ref{fig:energypaths}, which contains a subgraph of $\mathcal{G}$.  There are three vehicular routes, $r_1$, $r_2$, and $r_3$, and each passes through some of the nodes in the subgraph. Specifically, $r_1$ passes through Nodes 1 and 3; $r_2$ passes through Nodes 2, 3, and 4; and $r_3$ passes through Nodes 1, 2, 5, and 4. Suppose that we decide to transmit energy from Nodes 1 to 4. There are three energy paths in $\mathcal{P}(1,4)=\{p_1(1,4),p_2(1,4),p_3(1,4)\}$. $p_1(1,4)=\langle (1,3),(3,4)\rangle$ is constructed from two vehicular routes, where $(1,3)$ and $(3,4)$ come from $r_1$ and $r_2$, respectively. Hence, we have $p_1(1,4)=\langle r_1^1((1,3)),r_2^1((3,4))\rangle$.
$p_2(1,4)=\langle (1,2),(2,3),(3,4)\rangle$ is constructed from $r_3$ and $r_2$. We have $p_2(1,4)=\langle r_3^2((1,2)),r_2^2((2,3),(3,4))\rangle$.
Similarly, $p_3(1,4)=\langle (1,2),(2,5),(5,4)\rangle$ is constructed from $r_3$ only. Hence, we have $p_3=\langle r_3^3((1,2),(5,4))\rangle$.

In general, more nodes result in more edges in the network. The denser the network, the more energy paths we can establish between the energy sources and destinations. This results in higher flexibility in routing energy across the network. Consider the above example again. If we remove Node 3 and its associated arcs, there will be one energy path left connecting Nodes 1 and 4, i.e., $\mathcal{P}(1,4)=\{p_3(1,4)\}$. It can be easily seen that it is more flexible to route energy with three energy paths than one single path.  

The maximum amount of energy to be transmitted in each charging or discharging event is standardized and it can be carried by each EV in each charging-discharging cycle, i.e., the ``packet size'', denoted by $w$ units.\footnote{For the ease of modeling and implementation and without loss of generality, the same packet size of $w$ units is assumed for each charging-discharging cycle on each vehicle.} $w$ is a system parameter and its value should be set subject to the adopted WPT technology and equipment, the lengths of the (dis)charging facilities, and the battery capacities of the participating vehicles. It should be small enough so that the transmission of energy packet can complete in every charging and discharging event and that an appendance or removal of an energy packet from a vehicle will not make its state-of-charge fluctuating significantly. Let $f_i^j$ be the traffic flow of the $i$-th segment of $p_j$. By assigning some EVs along the sub-routes to carry energy, the energy transfer rate of $p_j$,
denoted by $g_j$ units, satisfies:
\begin{align} \label{eq:rate}
			g_j \leq w f_i^j, \quad i=1,\ldots,|p_j|.
\end{align}

\subsection{Charging-Discharging Cycle}
There is energy loss in both of the dynamic charging and discharging processes. The energy efficiencies of charging and discharging are given by $z_c$ and $z_d$, respectively, and thus, the fractions of $(1-z_c)$ and $(1-z_d)$ correspond to the portions of energy lost in charging and discharging, respectively. When a vehicle is employed to carry energy, there exists a charging-discharging cycle along each sub-route. For example, in Fig. \ref{fig:energypaths}, when some energy is decided to be transported from Nodes 1 to 3,  a vehicle can be charged along $r_1$ at Node 1 and then discharged at Node 3. Let $z=z_cz_d$. Hence, at the end of each charging-discharging cycle, only a fraction $z$ of energy can be retained. 

If the designated routes of all EVs are known, the EVs can be used to transmit energy more than one hop in $\mathcal{G}$ without discharging. Otherwise,
they need to be discharged and re-charged at each node (i.e., the charging-discharging cycle takes place at each hop). With the support of VANET, most vehicles are connected. Without loss of generality, the participating EVs are obligated to report (part of) their travel plans. In this way, for each route $r_i$, it can be told which EVs are traveling along the route.  
Suppose that the routes of all EVs are known. If $x$ units of energy are required to reach the destination of the energy path $p_j$, which is composed of $|p_j|$ sub-routes, $\frac{x}{z^{|p_j|}}$ units of energy need to be transmitted at the source of $p_j$. On the other hand, this incurs energy loss equal to $(\frac{1}{z^{|p_j|}}-1)x$ units.

\subsection{Delays}
Thanks to the opportunistic contacts of the vehicles via the (dis)charging facilities, VEN is not an instantaneous system. Time can become critical for energy transfer over VEN. In general, a delay characterizes how much time is required for a particular event to happen and we evaluate all time-related issues in terms of delays.
Due to its ``packet switching''-like characteristics, we can analyze the delay incurred to route energy packets from a particular source to a destination in terms of the following \cite{networking}:
\begin{itemize}
	\item Processing delay\\
	In computer networking, it refers to the time required to examine the packet's header and to determine the outgoing path at a router. While data in different data packets are basically unique, energy in VEN is a single commodity. As long as sufficient energy reaches the destination as planned, it is not important if the energy actually comes from the designated source or is ``borrowed'' from other sources. Therefore, there is basically no header in an energy packet. Furthermore, the processing delay may also be considered to capture the time required to transfer energy wirelessly between the vehicles and the charging facilities. As discussed in Section \ref{sec:VEN}, wireless energy transfer takes place when the vehicle is moving along its route. Energy transfer can be completed before the vehicle leaves the (dis)charging track (see Fig. \ref{fig:track}) if the packet size $w$ is sufficiently small or the tracks are sufficiently long. As vehicle travel time will be included in the propagation delay, processing delay can be ignored in VEN.
	\item Queuing delay\\
	As seen in Fig. \ref{fig:track}, wireless charging and discharging events normally take place simultaneously along the road segments installed with (dis)charging facilities. Moreover, it is not uncommon for a (dis)charging segment to serve multiple transmissions of different source-destination pairs at the same time. Together with the single commodity property, energy at the energy storage can be advanced to some extent that a charging event take place before a discharging event along an energy path. Therefore, energy packets do not need to be queued at the storage and thus there is no queuing delay.
	\item Transmission delay\\
	It refers to the amount of time needed to push all the energy packets onto an energy path. This depends on how many vehicles available traveling along the vehicular routes on the path. This is related to the energy transfer rate $g_j$ constrained by the traffic flow $f_i^j$ (See \eqref{eq:rate}).
	\item Propagation delay \\
	It refers to the time required to propagate an energy package from its source to its destination.
	At the (dis)charging facilities, energy can be transferred wirelessly from and to the vehicles on the move. 
It takes  time $d(p_j)$ units for energy to ``propagate'' along $p_j$, where
\begin{align} \label{propagation}
		d(p_j) = \sum_{i=1}^{|p_j|}d(r_i^j(n_i,m_i)) = \sum_{i=1}^{|p_j|}\sum_{a_k\in r_i^j(n_i,m_i)}d(a_k).
\end{align}
\end{itemize}

Therefore, the amount of time required to transport energy is composed of the total ``propagation delay'' and ``transmission delay'' along the transmission path. To transport $\frac{x}{z^{|p_j|}}$ units of energy along $p_j$ from its source, it takes a duration of $d(p_j)+\frac{x}{z^{|p_j|}g_j}$ units (see \eqref{propagation}). In other words, in a time window of $T$ units, the amount of energy transferable along $p_j$, denoted by $x_j$ units, is governed by $x_j\leq (T-d(p_j))z^{|p_j|}g_j,$ and the corresponding energy loss incurred is $(\frac{1}{z^{|p_j|}}-1)x_j$ units.
Therefore, the amount $x(s,t)$ units of energy transferred to Node $t$ from Node $s$ in a time period $T$ satisfies:
\begin{align}
x(s,t) = \sum_{j|p_j\in \mathcal{P}(s,t)}{x_j} \leq 
\sum_{j|p_j\in \mathcal{P}(s,t)}{(T-d(p_j))z^{|p_j|}g_j}
\label{transferredEnergy}
\end{align}
The corresponding amount of energy loss of $L(s,t)$ units is given by:
\begin{align}
	L(s,t) = \sum_{j|p_j\in \mathcal{P}(s,t)}{(\frac{1}{z^{|p_j|}}-1)   x_j}.
	\label{energyLoss}
\end{align}

%---------------------------------------------------------------------------
\section{System Analysis} \label{sec:formulation}

To utilize VEN, the system needs to be configured by setting up the appropriate energy paths, each of which can deliver enough energy within a given time frame. Two key concerns are the transferable amount of energy and the energy loss. They will be studied as optimization problems and give some analytical results.

A small packet size of $w$ units is sufficient to facilitate a noticeably large amount of energy transfer over a large geographical area and this will be verified in Section \ref{sec:simulation}. Since a practical value of $w$ is very small when compared with the battery capacity of typical EVs, for a single-battery EV, only a small portion of the battery will be reserved for VEN while the rest can still be used to support normal EV operations. It can basically be assumed that an EV can serve multiple energy paths by carrying multiple energy packets simultaneously. 
It is unlikely that an EV cannot serve due to battery overflow. As discussed, $w$ should be carefully chosen with a sufficiently small value by considering the battery sizes of all participating EVs. Moreover, the longer a vehicle runs, the more energy it consumes and the more room available for it to store energy for VEN in its battery. Furthermore, as pointed out in \cite{VEN_conf}, there exists an option of installing a secondary battery in a participating EV dedicated to carrying energy in VEN. In this case, the variation of energy-carrying capabilities due to different EV as original battery capacities can be eliminated.

We first study how to convey energy over the energy paths by maximizing the total amount of transferred energy subject to a maximum tolerable energy loss.
Given the source-destination pair $(s,t)$, the set of energy paths $\mathcal{P}(s,t)$ can be determined based on the method discussed in \cite{VEN_infocom}. Suppose that we have an energy loss requirement with an upper limit $\overline{L}$ (called the maximum tolerable energy loss). Based on Section \ref{sec:model}, the problem is formulated as follows:
\begin{problem}[Energy Transfer Maximization]
\label{maxopt}
\begin{subequations}
\label{maxopteq}
\begin{align}
\text{maximize}\quad 	& \sum_{j=1}^{|\mathcal{P}(s,t)|}{x_j} \label{5a}\\
\text{subject to}\quad 
& x_j\leq (T-d(p_j))z^{|p_j|}g_j, \quad j=1,\ldots, |\mathcal{P}(s,t)|, \label{5b}\\
& g_j \leq w f_i^j, \quad i=1,\ldots,|p_j|, j=1,\ldots, |\mathcal{P}(s,t)|, \label{5c}\\
& \sum_{j=1}^{|\mathcal{P}(s,t)|}{(\frac{1}{z^{|p_j|}}-1)   x_j} \leq \overline{L} \label{5d}\\
& x_j\geq 0, \quad j=1,\ldots, |\mathcal{P}(s,t)| \label{5e}.
\end{align}
\end{subequations}
\end{problem}
The total energy transferred is maximized in \eqref{5a}, subject to the constraint of the transferred amount on each path \eqref{5b}, the energy transfer rate constraint \eqref{5c}, the energy loss constraint \eqref{5d}, and the non-negativity constraint \eqref{5e}. Given $\mathcal{P}(s,t)$, $|\mathcal{P}(s,t)|$, $d(p_j)$'s and $|p_j|$'s are constants. $z$, $w$, $T$, $f_i^j$'s, and $\overline{L}$ are system parameters while $x_j$'s and $g_j$'s are variables. It can be seen that Problem 1 is indeed a linear program (LP) and the solution can be easily computed with a standard LP solver.

In some cases, we may just need to convey a given amount of energy and retain the flexibility on the energy loss. The total energy loss incurred can be minimized provided that a given minimum energy amount $\underline{X}$ (also called the minimum required transferable energy) can be achieved. The problem is formulated in the following:
\begin{problem}[Energy Loss Minimization]
\label{minopt}
\begin{subequations}
\label{minopteq}
\begin{align}
\text{minimize}\quad 	& \sum_{j=1}^{|\mathcal{P}(s,t)|}{(\frac{1}{z^{|p_j|}}-1)   x_j} \label{6a}\\
\text{subject to}\quad 
& x_j\leq (T-d(p_j))z^{|p_j|}g_j, \quad j=1,\ldots, |\mathcal{P}(s,t)| \label{6b}\\
& g_j \leq w f_i^j, \quad i=1,\ldots,|p_j|, j=1,\ldots, |\mathcal{P}(s,t)| \label{6c}\\
& \sum_{j=1}^{|\mathcal{P}(s,t)|}{x_j} \geq \underline{X}  \label{6d}\\
& x_j\geq 0, \quad j=1,\ldots, |\mathcal{P}(s,t)| \label{6e}.
\end{align}
\end{subequations}
\end{problem}
The total energy loss is minimized in \eqref{6a}, subject to the energy transfer constraint \eqref{6d} and some similar constraints as in Problem 1, where $\underline{X}$ is a system parameter. Similar to Problem 1, Problem 2 is also an LP and the solution can be easily determined. Once the energy paths have been constructed, the problems are relatively easy to solve. Note that, if it happens that an EV cannot serve temporarily due to whatever reason (e.g., with battery overflow), we may not allow that EV from further carrying more energy packets until some energy has been discharged. We can always do so as the vehicles are connected and can be tracked. If needed, we may adjust the traffic flow $f_i^j$ to reflect the situation and resolve Problem 1 or 2, again. The re-computation will not induce serious problems as they are just LP and can be solved very effectively.

Instead of showing how the problems are solved with numerical examples, some analytical results is given below to obtain more insights.
Consider that both energy transfer maximization and energy loss loss minimization need to be implemented simultaneously. Given the maximum tolerable energy loss $\overline{L}$ and the minimum required transferable energy $\underline{X}$, the following bi-objective optimization can be constructed:
\begin{problem}[Bi-objective Optimization]
\label{ob-obj}
\begin{subequations}
\label{biopt}
\begin{align}
\text{minimize}\quad 	& [-\sum_{j=1}^{|\mathcal{P}(s,t)|}{x_j},\sum_{j=1}^{|\mathcal{P}(s,t)|}{(\frac{1}{z^{|p_j|}}-1)\tilde{x}_j}] \label{bi_obj}\\
\text{subject to}\quad 
& \sum_{j=1}^{|\mathcal{P}(s,t)|}{x_j} \geq \underline{X}  \label{bi_a}\\
& \sum_{j=1}^{|\mathcal{P}(s,t)|}{(\frac{1}{z^{|p_j|}}-1)   x_j} \leq \overline{L} \label{bi_d}\\
& x_j\leq (T-d(p_j))z^{|p_j|}g_j, \quad j=1,\ldots, |\mathcal{P}(s,t)|, \label{bi_b}\\
& g_j \leq w f_i^j, \quad i=1,\ldots,|p_j|, j=1,\ldots, |\mathcal{P}(s,t)|, \label{bi_c}\\
& x_j\geq 0, \quad j=1,\ldots, |\mathcal{P}(s,t)| \label{bie}.
\end{align}
\end{subequations}
\end{problem}

\begin{theorem} \label{thm:XL}
Suppose
\begin{align} 
z^{|p_j|}\leq \frac{1}{2}, j=1,\ldots, |\mathcal{P}(s,t)|. \label{thmCondition}
\end{align}
We have the following results:
\begin{enumerate}
	\item $\underline{X}\leq\overline{L}$ is a necessary condition for the feasibility of Problem \ref{ob-obj}.
	\item Let $f_1^*(x)$ be the optimal objective function value of Problem \ref{maxopt}. When $\underline{X}$ is set larger than $f_1^*(x)$, Problem \ref{ob-obj} is infeasible.
	\item Let $f_2^*(x)$ be the optimal objective function value of Problem \ref{minopt}. When $\overline{L}$ is set smaller than $f_2^*(x)$, Problem \ref{ob-obj} is infeasible.
\end{enumerate}
\end{theorem}

\begin{proof}
\begin{enumerate}
	\item By \eqref{thmCondition}, \eqref{bi_a}, and \eqref{bi_d}, we have
		\begin{align}
			\underline{X}	&	\leq \sum_{j=1}^{|\mathcal{P}(s,t)|}{x_j} 
							\leq \sum_{j=1}^{|\mathcal{P}(s,t)|}{(\frac{1}{z^{|p_j|}}-1)x_j}
							\leq \overline{L}. \label{thm:ineq}
		\end{align}
		If $\underline{X}>\overline{L}$, there will not exist an $\tilde{x}$ which can satisfy \eqref{thm:ineq}. Hence Problem \ref{ob-obj} is infeasible unless $\underline{X}\leq\overline{L}$.
		\item Let $x=[x_1,\ldots,x_{|\mathcal{P}(s,t)|}]^T$, $f_1(x)=\sum_{j=1}^{|\mathcal{P}(s,t)|}{x_j}$, and $f_2(x)=\sum_{j=1}^{|\mathcal{P}(s,t)|}{(\frac{1}{z^{|p_j|}}-1)x_j}$. Suppose that Problem \ref{ob-obj} is feasible.  It has a set of optimal solutions $\tilde{\mathcal{X}}=\{\tilde{x}\}$ constituting a Perato frontier $\{f_1(\tilde{x}),f_2(\tilde{x})\}$ such that $\underline{\delta}_1\leq f_1(\tilde{x})\leq \overline{\delta}_1$ and $\underline{\delta}_2\leq f_2(\tilde{x})\leq \overline{\delta}_2, \forall \tilde{x}\in\tilde{\mathcal{X}}$, where $\underline{\delta}_1=\inf\{f_1(\tilde{x})\}$, $\overline{\delta}_1=\sup\{f_1(\tilde{x})\}$, $\underline{\delta}_2=\inf\{f_2(\tilde{x})\}$, and $\overline{\delta}_2=\sup\{f_2(\tilde{x})\}$. By introducing the $\underline{X}$ from Problem \ref{ob-obj}, Problem \ref{maxopt} can be re-written as 
		\begin{subequations}
			\label{maxopt2}
			\begin{align}
			\text{minimize}\quad 	& -f_1(x) \\
			\text{subject to}\quad 
			& \sum_{j=1}^{|\mathcal{P}(s,t)|}{x_j} \geq \underline{X} \label{max2a}\\
			& \eqref{5b}, \eqref{5c}, \eqref{5d}, \eqref{5e}. 
			\end{align}
		\end{subequations}
		Since Problem \ref{ob-obj} is feasible, \eqref{bi_a} holds, so as \eqref{max2a}. Then \eqref{maxopt2} is equivalent to \eqref{maxopteq}. Furthermore, with the same feasible region, solving \eqref{maxopt2} is as solving \eqref{biopt} without $f_2(x)$ and this gives $f_1^*(x)=\overline{\delta}_1$ for maximization. However, $\underline{X}>f_1^*(x)$ violates \eqref{thm:ineq}, which gives an contradiction.
		\item Similar to the proof in Part (2) above, by introducing the $\overline{L}$ from Problem \ref{ob-obj}, Problem \ref{minopt} can be re-written as 
		\begin{subequations}
		\label{minopt2}
		\begin{align}
		\text{minimize}\quad 	& f_2(x)\\
		\text{subject to}\quad 
		& \sum_{j=1}^{|\mathcal{P}(s,t)|}{(\frac{1}{z^{|p_j|}}-1)   x_j} \leq \overline{L} \label{min2a}\\
		& \eqref{6b}, \eqref{6c}, \eqref{6d}, \eqref{6e}.
		\end{align}
		\end{subequations}	
			\eqref{minopt2} is equivalent to \eqref{minopteq}. With the same feasible region, solving \eqref{minopt2} is as solving \eqref{biopt} without $f_1(x)$ and this gives $f_2^*(x)=\underline{\delta}_2$ for minimization. However, $\overline{L}<f_2^*(x)$ violates \eqref{thm:ineq}, which gives an contradiction.
\end{enumerate}
\end{proof}
Condition \eqref{thmCondition} is met when the energy efficiency $z$ is low or the energy paths are composed of many vehicular sub-routes. The latter may be due to the source and destination being too far apart or limited vehicular routing information.
Theorem \ref{thm:XL} is useful when the total transferred energy and energy loss need to be optimized simultaneously. 
Solving a simpler problem, i.e., Problem \ref{maxopt} or \ref{minopt}, gives us ideas on how to set the values of the parameters, $\underline{X}$ and $\overline{L}$, for the harder  Problem \ref{ob-obj}.

%---------------------------------------------------------------------------
\section{Economic Model} \label{sec:economic}
\begin{figure}[!t]
\centering
\includegraphics[width=3.2in]{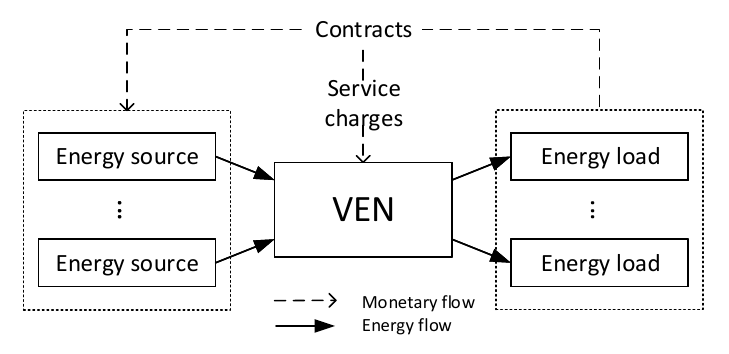}\vspace{-0.5cm}
\caption{Role of VEN between the energy sources and loads.}
\label{fig:revenue}
\end{figure}

VEN is designed as an energy distribution system, in which both the sources and loads of energy are not parts of the system (see Fig. \ref{fig:revenue}). It provides energy delivery services for the users, i.e., energy source-load pairs, and the service charge may depend on many factors, e.g., distances between sources and loads, delivery time frames, amounts of deliverable energy, quality of service, etc. The detailed service charging model depends on the applications and it is out of the scope of this work. In order to assess the economic feasibility, a simple economic model is developed for renewable energy dispatch, which will be used to evaluate the system performance in Section \ref{sec:simulation}. In this application, the ``offline'' renewable energy is delivered, due to the excessive generation which cannot be brought to the grid, to appropriate destinations over VEN.

The economic value of VEN is evaluated by the annual revenue ($R_{ann}$) minus the annual cost ($C_{ann}$). Suppose that the revenue comes from the annual amount of energy delivered at the destinations, i.e.,
\begin{align*}
R_{ann} = R_{e}E_{d} = R_{e}E_{s} \epsilon,
\end{align*}
where $R_{e}$, $E_{d}$, $E_{s}$, and $\epsilon$ stand for electricity rate, annual delivered and generated amounts of energy, and overall system efficiency, respectively.

As usually only a portion of renewable generation can be brought ``online'' and utilized in the grid, it is considered that VEN can improve the utilization of renewables by manipulating the ``offline'' energy. Thus it is assumed that the ``offline'' renewable energy is free. Then the cost mainly stems from the energy storage for temporary repository ($C_{s}$), the (dis)charging facilities ($C_{f}$), and EVs ($C_{v}$). 
$C_{s}$ is estimated based on the levelized cost of energy (LCOE) \cite{storagecost}, which is the ratio of lifetime costs to lifetime delivered energy. It takes into account factors such as capital cost, financing cost, fixed and variable operations and maintenance cost, installation cost (components and labor), etc. As seen from Fig. \ref{fig:track}, the storage manages temporary energy imbalance among the passing-by EVs. Total amount of energy delivered by the storage depends on the real dynamics on the roads. This amount is modeled as a fraction $\delta_s$ of total generated energy and thus we have $C_s = C^{\text{LCOE}}_{s}E_s\delta_s$, where $C^{\text{LCOE}}_{s}$ is LCOE of the storage.
 
For the dynamic (dis)charging facilities, the technology is still in the research and development stage and thus its levelized cost is not available. The capital cost of such a facility is estimated by considering a similar technology for dynamic charging reported in \cite{EconDCEV}. The cost model given in \cite{EconDCEV} is adopted to estimate the lifetime cost of (dis)charging facilities, which is then annualized by multiplying with the capital recovery factor ($R_{CRF}$) \cite{V2Grevenue}, which is defined as $R_{CRF}=\frac{\tau_f}{1-(1+\tau_f)^{-\eta_f}}$, where $\tau_f$ and $\eta_f$ are the discount rate of the equipment and the number of years the equipment lasts, respectively. Hence we have $C_{f}= (c^f_f+c^{v}_fl_{f})n_{f}R_{CRF}$, where $c^f_f$, $c^v_f$, $l_{f}$, and  $n_f$, are the fixed cost, the per-unit length variable cost, average track length, and the required quantity, respectively. 

In VEN, EVs are responsible for transporting energy but auxiliary in nature; they are generally not owned by the system but incentivized to carry additional energy for the system on their own. It is simply assumed that a part of the revenue is used as the incentive and thus we have $C_v = R_v R_{ann}$, where $R_v$ is the incentive rate. Therefore the total annual cost is given by

\begin{align*}
\vspace{-0.9cm}
	C_{ann} &=  C_{s} + C_{f} + C_{v} \\
	&=  C^{\text{LCOE}}_{s}E_{s}\delta_s + (c^f_f+c^v_{f}l_{f})n_{f}R_{CRF} + R_v R_{ann}.
\end{align*}

\section{Feasibility Study} \label{sec:simulation}

In this section, the operational and economic feasibilities of VEN are evaluated. Wind energy dispatch is selected as an application example of VEN. For the former,  it is demonstrated that a significant amount of energy can be transmitted through VEN in a reasonable period of time. For the latter, the economic model discussed in Section \ref{sec:economic} is evaluated for wind energy dispatch.

\subsection{Operational Feasibility} \label{subsec:operational}
Consider that the renewable energy produced from the wind farms located in remote areas of the United Kingdom (U.K.) is transmitted to the city of London.  A VEN is constructed by adopting the existing U.K. road networks.  Based on \cite{ukroads}, the network is created with 998 nodes and 2470 arcs ($a_i$'s). One set of real traffic data acquired in June 2013 is adopted to specify the journey time, distance, and vehicular flow for each of the road segments (i.e., arcs). 4788 vehicular routes ($r_i$'s) is randomly created in the road network, each of  which is composed of several connected road segments  no longer than 200 km. The traffic flow and total journey time of each $r_i$ are set with the minimum of vehicular flows of the corresponding composite $a_i$'s and the sum of individual journey times of the composite $a_i$'s, respectively.

\begin{figure}[!t]
\centering
\includegraphics[width=2.7in]{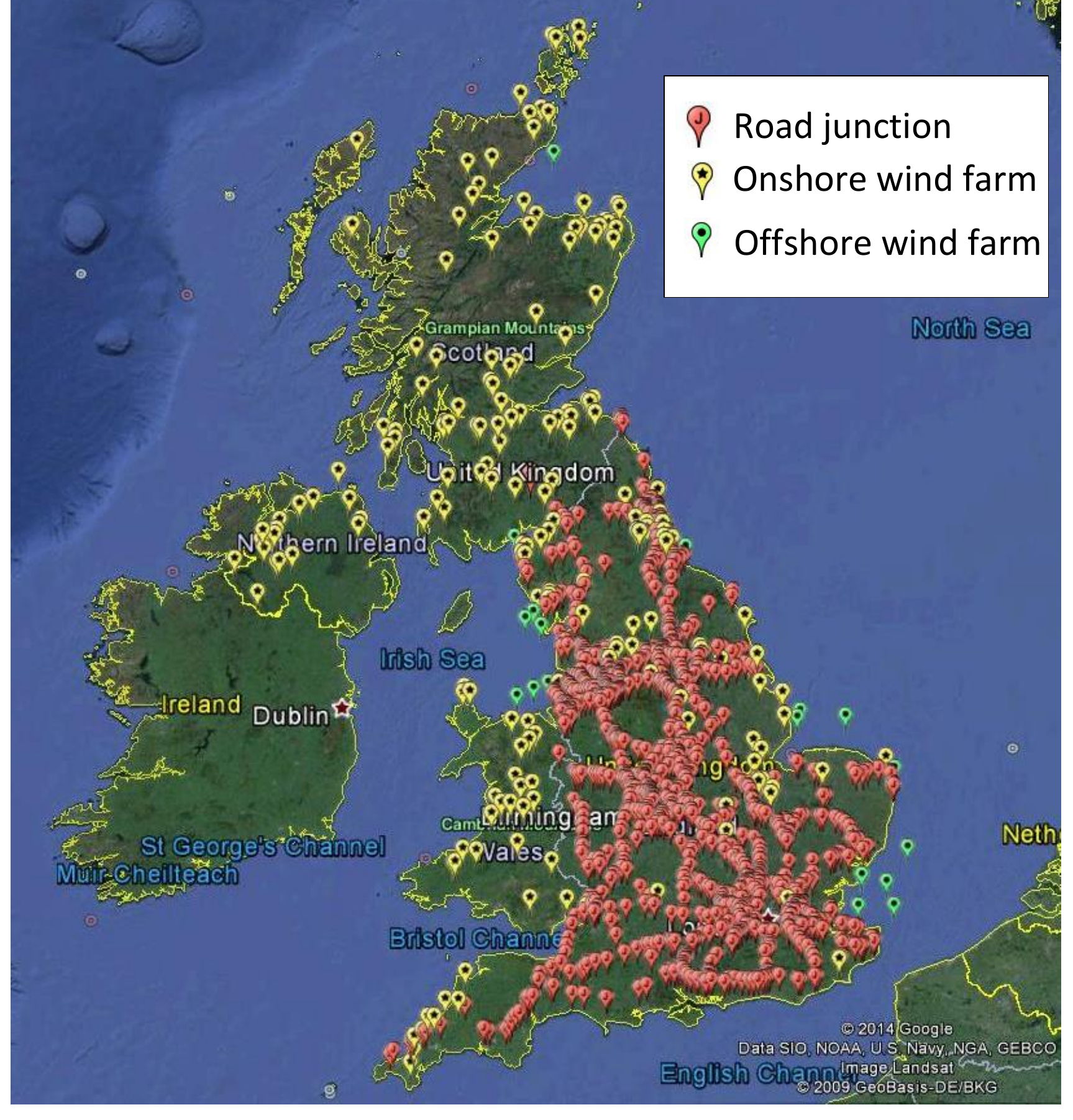} \vspace{-0.3cm}
\caption{Locations of road junctions, and onshore and offshore wind farms (adopted from Google Earth).}
\label{fig:UKmap}
\vspace{-0.2cm}
\end{figure}

U.K. is the sixth largest nations producing wind energy, with an annual energy
production of more than $26\times 10^6$ MWh \cite{UKWindDB}, but only 26\% was brought online in 2013 \cite{UKWind}. According to \cite{offshore} and \cite{onshore}, there are 203 onshore and 20 offshore wind farms. Fig. \ref{fig:UKmap} shows the locations of the road junctions and the wind farms. 67 nodes are selected in the the road network, which are close to the wind farms, as the energy sources ($s$'s) in VEN. One junction near London is selected as the energy destination $t$ for demonstrative purpose. Energy paths $p_j(s,t)$'s are constructed by augmenting sub-routes of the 4788 $r_i$'s. Energy is transmitted from the sources to the destination along \textit{some} of the energy paths.

Consider the nominal setting of  0.1\% EV penetration rate,\footnote{The EV penetration rate refers to the percentage of participating EVs in the vehicle population. When the EV penetration rate is 0.1\%, one out of 1000 cars supports VEN.} packet size $w$ of $0.1$ kWh\footnote{An existing dynamic charging prototype, OLEV, can support charging rate of 100 kW \cite{kaist}, and an energy packet of 0.1 kWh will take 3.6 s. Thus the packet size of 0.1 kWh is feasible.}, efficiency rate $z$ of $0.9$\footnote{The industry expects that the efficiency of wireless energy transfer on moving vehicles can exceed 90\% \cite{theengineer}. NASA has developed a prototype called EVWireless with over 90\% energy efficiency \cite{EVWireless}.}, and time period $T$ of 5 hours.  As most popular EVs available in the market are equipped with batteries of more than 20 kWh \cite{batterycap}, our energy packet size is rather conservative. In the following, the total transferred energy (reaching London from the wind farms) and the corresponding energy loss are studied by varying one parameter in each case.  Each data point is computed from the right-hand side of either \eqref{transferredEnergy} or \eqref{energyLoss}, in which, for simplicity, only some paths are considered instead of all possible energy paths.

\begin{figure*}[!t]
	\begin{center}
		\subfigure[At different efficiencies.]{\label{fig:changez}\includegraphics[width=3.00in]{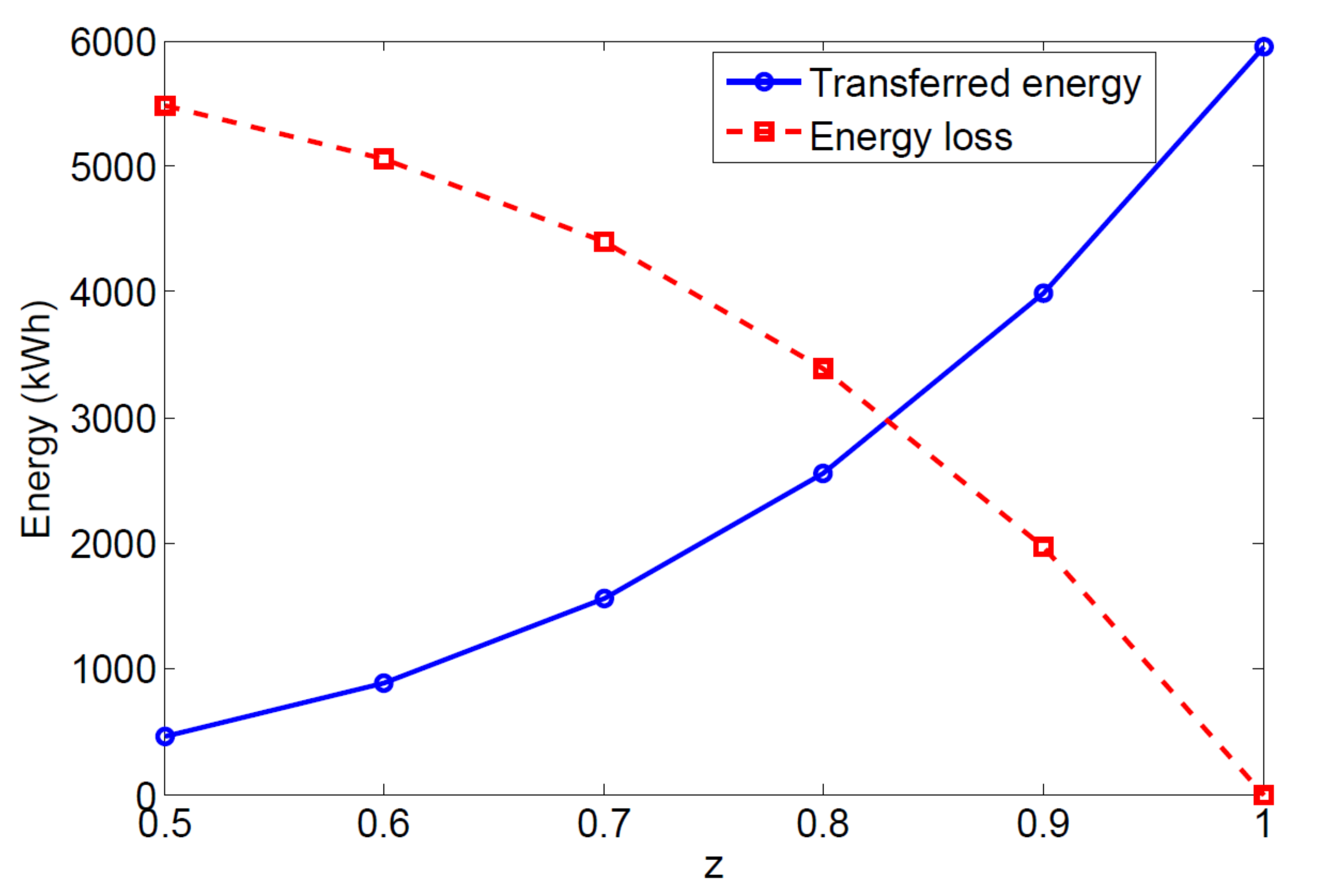}}
    	\subfigure[In various periods of time.]{\label{fig:changeT}\includegraphics[width=3.00in]{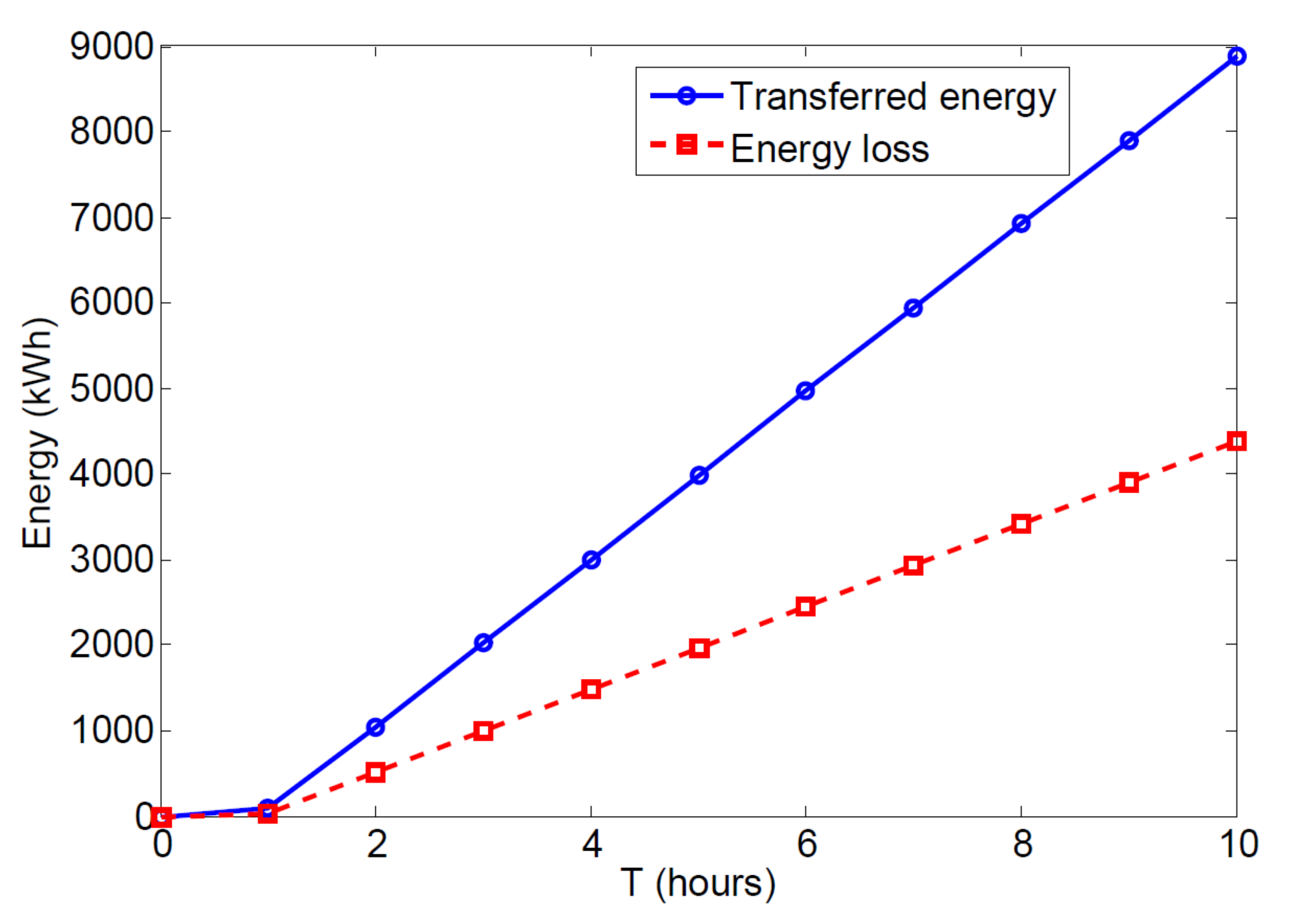}} 
    	\subfigure[With various buffer sizes.]{\label{fig:changew}\includegraphics[width=3.00in]{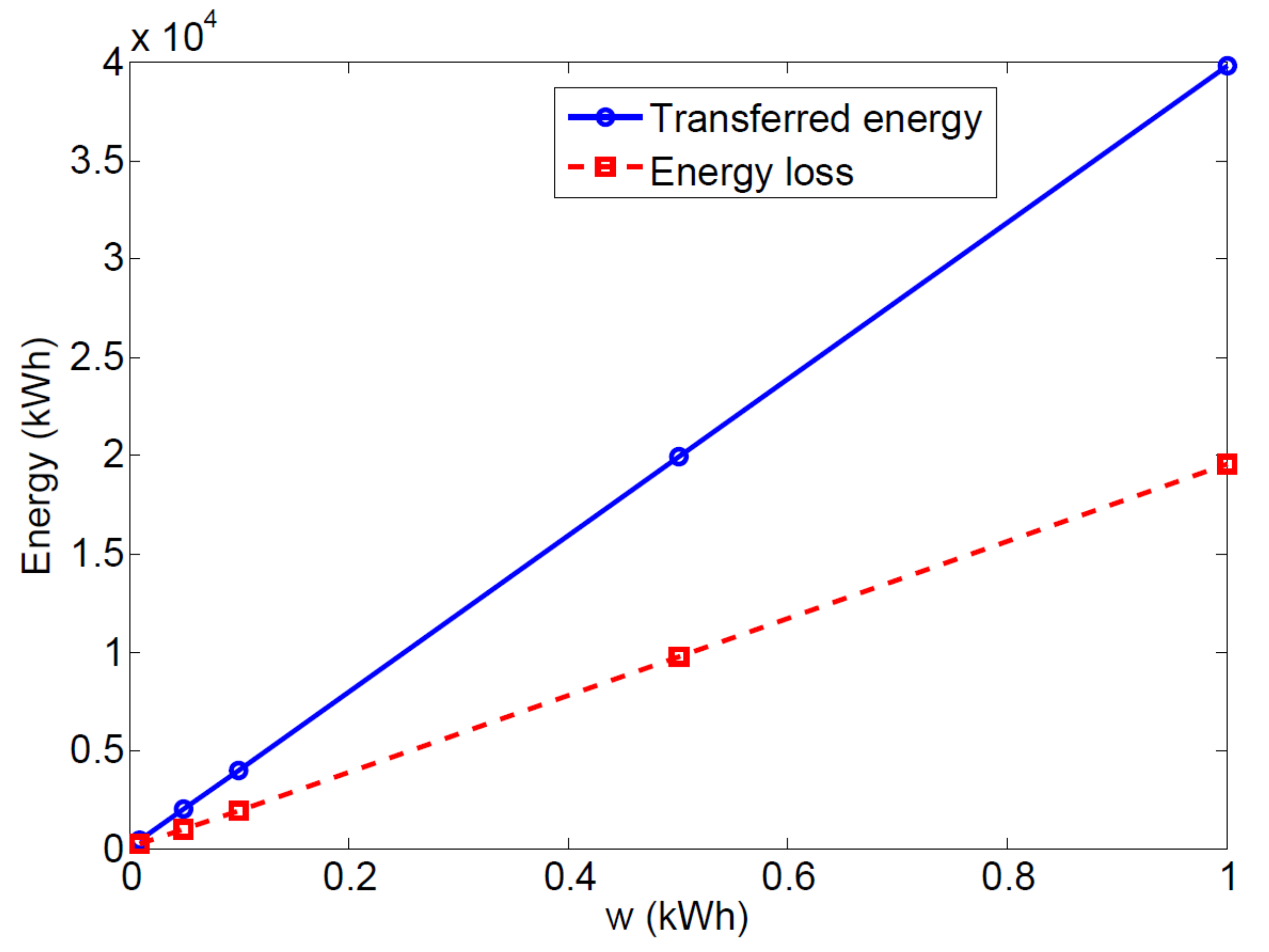}}
		\subfigure[With different EV penetration rates.]{\label{fig:changePen}\includegraphics[width=3.00in]{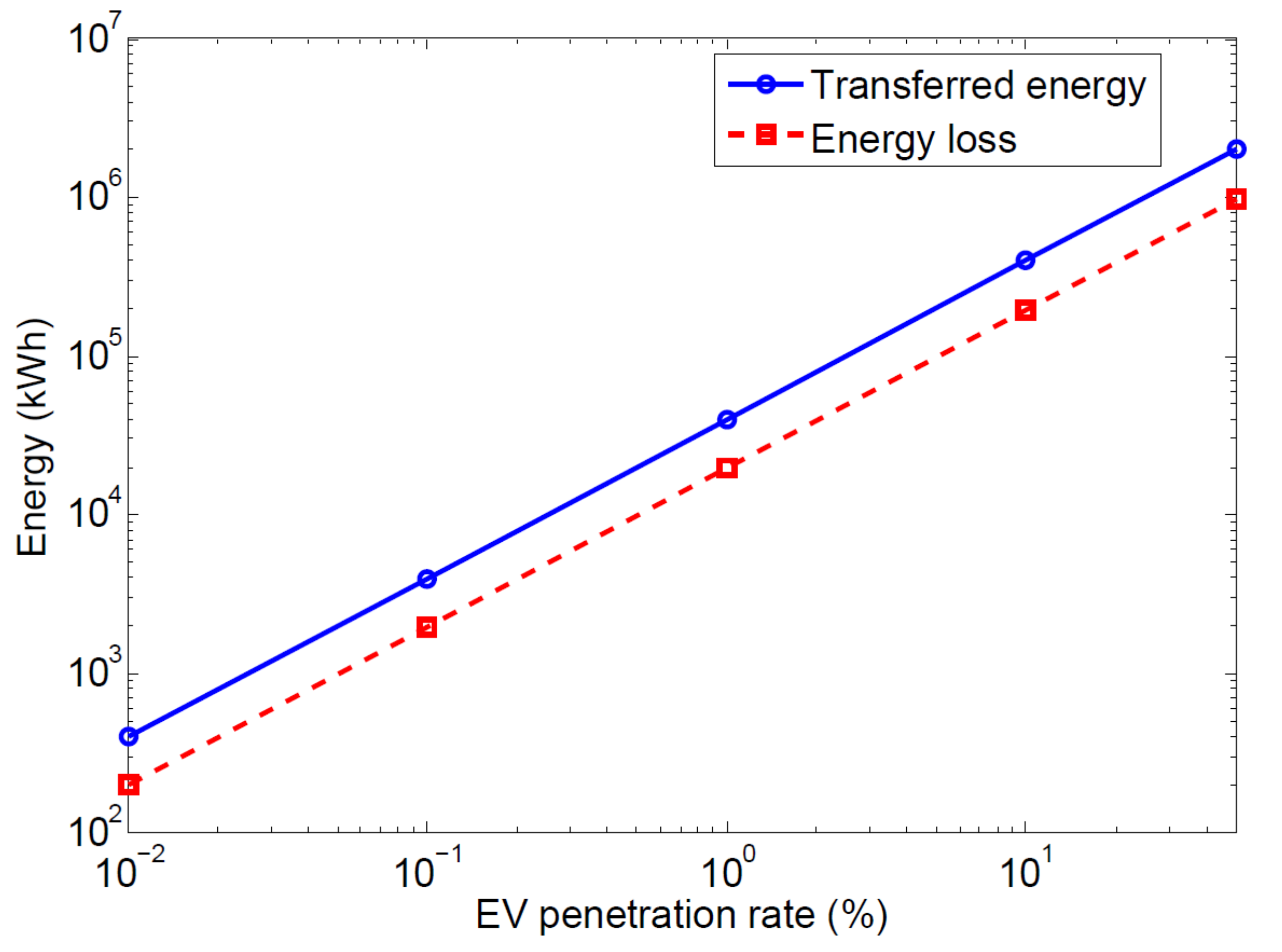}}
	\end{center}
	\caption{Transferred energy and energy loss.}
  \label{fig:sim}
\end{figure*}

Fig. \ref{fig:changez} shows the impact of total transferred energy and energy loss by changing $z$ from the nominal setting. In five hours, the amounts of energy transferred are in the order of MWh. Note that the allowed period has already included the time required for the EVs to move from one place to another. The more efficient the wireless energy transfer technology, the more energy can be transferred with smaller energy loss.
As an extreme case, taking $z$ as one gives no energy loss. This can be achieved only when the system is perfect; the energy efficiency is $100\%$. It is just a speculative result.
When $z$ is smaller than 0.83,\footnote{This threshold actually depends on the energy paths chosen for energy transmission.} the amount of energy loss can overshoot the transferred energy. Hence $z$ is an important factor when VEN is designed. However, if there is no VEN or other similar energy transmission architecture available, the huge amount of renewable energy may need to be stored at the original generation locations and eventually wasted due to surplus renewable production. As the technology of wireless energy transfer advances, the overall efficiency will improve.
Fig. \ref{fig:changeT} studies the impact of the allowed time period. With $T
\leq 1$ hour, the transferred energy is very small as most of the EVs carrying the energy have not  reached the destination. With a larger $T$, more transferred energy and energy loss occur. The gap between transferred energy and energy loss grows with $T$ showing that the energy transfer is more efficient in a longer time period.
Fig. \ref{fig:changew} illustrates the impact of the packet size. More energy will reach the destination and more energy is  also lost when a larger amount of energy can be (dis)charged each time. A larger packet size leads to a higher efficiency as more energy can be transmitted than  lost. 
Finally, Fig. \ref{fig:changePen} shows that the transferred energy and energy loss grow linearly with the EV penetration rate. The penetration rate affects the number of EVs providing the service only. The more EVs available, the more energy can be transferred. At the same time, the more frequent charging and discharging events result in higher energy loss. So both lines have similar slopes.

To summarize, a VEN is proposed to transport renewable energy from remote wind farms to London. Note that the total transferable energy and loss shown above are not optimal, as \textit{some} of the energy paths are just selected to transport the energy. In fact, there may be more efficient paths experiencing fewer charging-discharging cycles. Nevertheless, the amount of energy transferable through VEN in our illustration is still very significant. Although there is energy loss in VEN, most of the  energy to be transported in VEN would be lost anyway without VEN. 
This shows that VEN can complement the power network and enhance the overall power transmission rate.
Besides pure energy transfer, there exist other applications in which an energy source and destination do not have direct energy transfer relationship. We may also consider VEN as an energy storage system spanning a large geographical area with many interfaces (i.e., nodes in VEN). It can be used to absorb energy excessively generated at the ``energy sources'' and release energy at the ``energy destinations'' where energy is in deficit. Moreover, we may also use the energy ``stored'' in VEN to charge EVs for their energy consumption or for other purposes. This implicitly creates more loads in the network and may improve the renewable energy utilization.

\vspace{-0.3cm}
\subsection{Economic Feasibility}

\begin{table}[!t]
\renewcommand{\arraystretch}{1.3}
\caption{Parameter settings.}
\label{tab:parameters}
\centering
\scriptsize
\begin{tabular}{l|c|l}
\hline\hline
Parameter & Value & Remarks\\
\hline
$E_s$ 			&  $28.1\times 0.7$ TWh		& \cite{UKWind}  \\
$\epsilon$				&  0.67	 & Estimated from Fig. \ref{fig:sim} 	\\
$R_e$ 	&  0.154	USD/kWh	& Tariff of U.K. \cite{tariff}		\\
$C_{s}^{LCOE}$ 		& 0.51 USD/kWh & Converted from 45.1 c\euro/kWh \cite{storagecost}\\
$c_f^f$						& 50000 USD		& \cite{EconDCEV}\\
$c_f^v$						& 500 USD/m	& \cite{EconDCEV}\\
$l_f$					&  1000 m		\\
$\tau_f$			&	0.1		\\
$\eta_f$ 			&	 10	years	\\
$n_f$								& 998		\\
$R_v$								&  0.1	\\
\hline\hline
\end{tabular}
\vspace{-0.3cm}
\end{table}

\begin{figure}[!t]
	\begin{center}
		\subfigure[At various equipment discount rates.]{\label{fig:econ1}\includegraphics[width=2.9in]{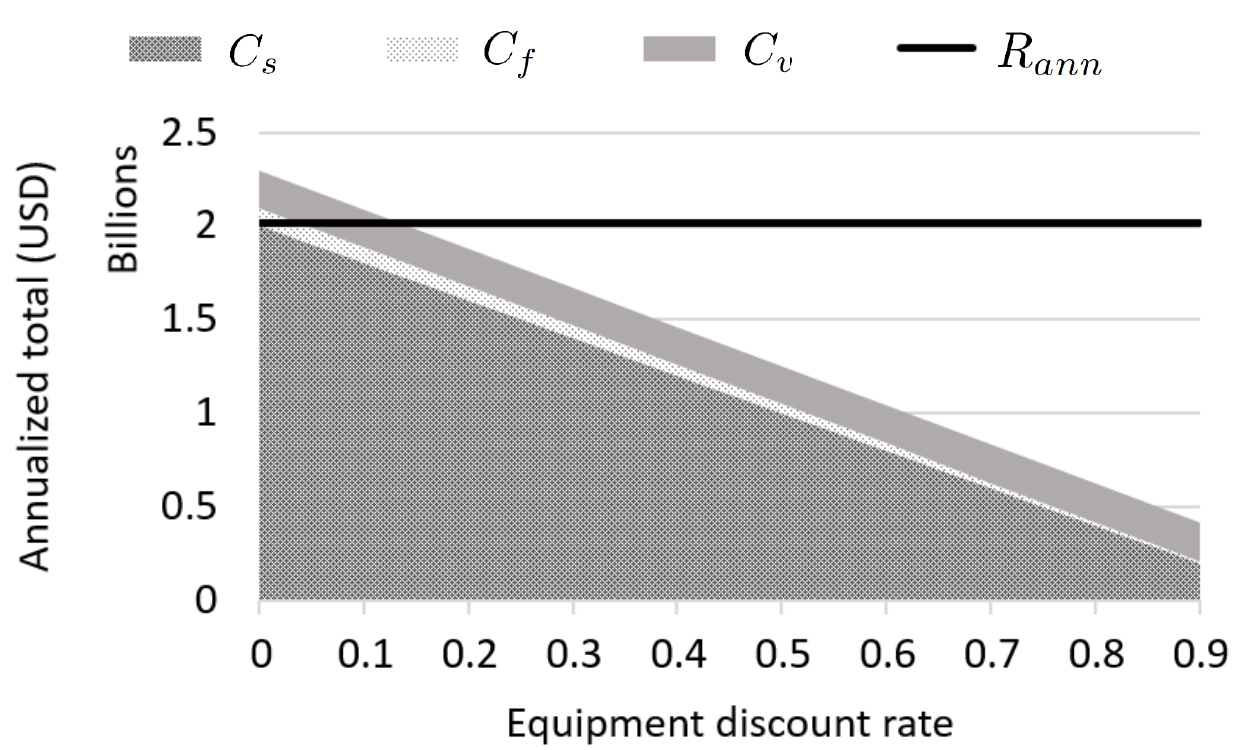}}
    	\subfigure[At various energy amounts in storage.]{\label{fig:econ2}\includegraphics[width=2.9in]{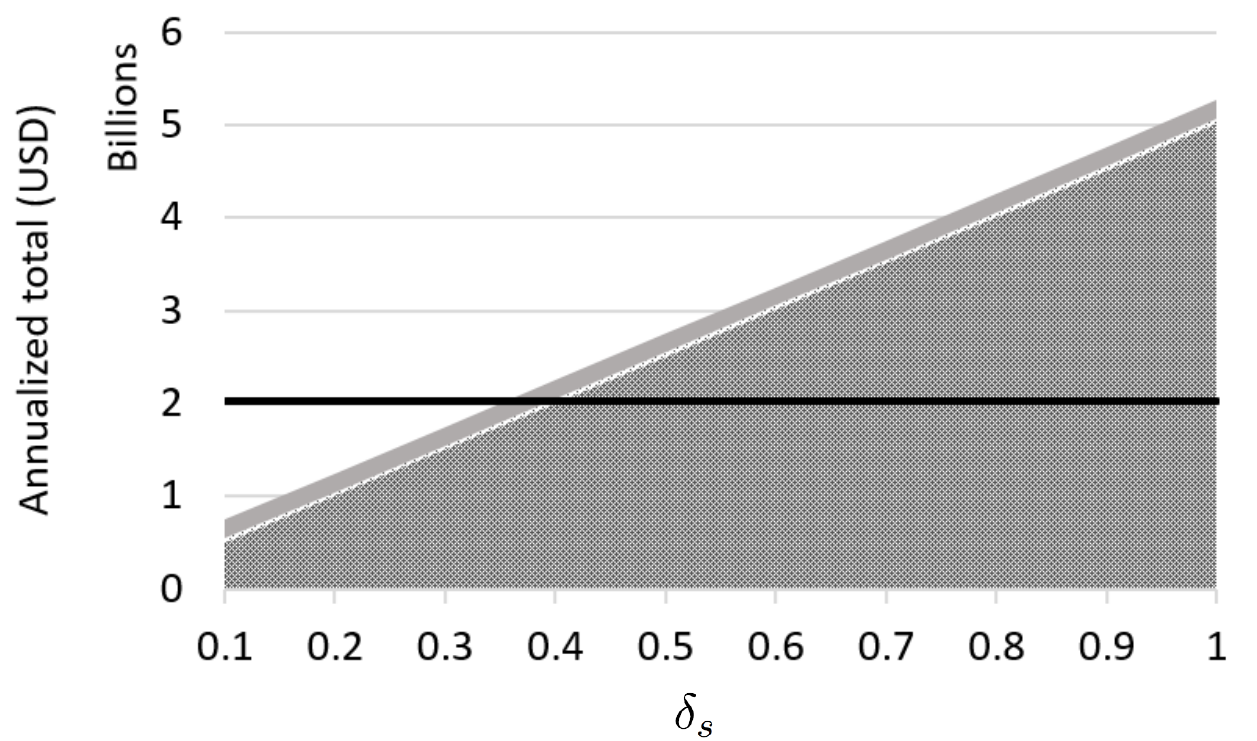}} 
	\end{center}
	\caption{Total profit.}
  \label{fig:econ}
	\vspace{-0.6cm}
\end{figure}

The total possible profit is estimated when implementing wind energy dispatch illustrated in Section \ref{subsec:operational} for a year. The parameter settings are given in Table \ref{tab:parameters}. U.K. generated 28.1 TWh of wind energy in 2014 \cite{UKWind} and it is assumed that 70\% of the generation is delivered through VEN (after excluding the online wind power directly transmitted to the grid). As observed in Fig. \ref{fig:sim}, around $2/3$ of the total generated energy can reach the destination and thus it is assumed that the overall system efficiency equals $0.67$.\footnote{The actual system efficiency highly depends on the settings of the real situation, including locations of the sources and destinations, network structure, traffic, and energy routing. It cannot be easily quantified and thus the value is estimated from our study.} 
As discussed in Section \ref{sec:VEN}, the energy storage has relatively short (dis)charging time and the required capacity is generally small. Supercapacitor is considered the best choice among all kinds of battery technologies for the energy distribution purpose and its LCOE is 0.51 USD/kWh \cite{storagecost}. The cost of the (dis)charging facility is estimated based on the prototype given in \cite{EconDCEV} and the (dis)charging track is assumed to be installed at each road junction is 1000 m long. The discount rate for the capital recovery factor is set to 0.1 and each piece of equipment is assumed to last for 10 years. Consider that one (dis)charging facility is installed at each of the 998 road junctions and 10\% of the revenue is allocated as the incentive cost for EVs.\footnote{The amount is estimated on an annual basis. Each EV may contribute differently during the year. How the incentive cost is distributed depends on the business model, which is beyond the scope of this paper.} Fig. \ref{fig:econ1} depicts the annualized total cost and revenue of the wind energy dispatch at different equipment discount rate, which accounts for how much discount of the cost when the relevant technologies advance and the equipment can be massively produced in the future. It can be seen that storage represents most of the cost. Without any discount, the system is not economical to operate. When the equipment cost drops more than 15\%, the system becomes profitable. 
As forecast by the World Energy Council, there will be 70\% drop in energy storage costs by 2030 \cite{worldenergy} and it is not hard to imagine that VEN will become economically feasible when the relevant technologies become mature.
Moreover, as the amount of energy stored during operation depends on the EV conditions at the road junctions, $\delta_s$ is a rough estimate. So $\delta_s$ is varied and  how $\delta_s$ affects the total profit is investigated. To account for future technological advances, The various equipment costs is considered to be dropped by half. Fig. \ref{fig:econ2} shows the annualized costs and total revenue corresponding to different $\delta_s$. The total cost increases with $\delta_s$ and the system becomes economically infeasible when $\delta_s$ gets large. This suggests that the system can produce higher profits when the EVs go in platoons such that charging and discharging can take place simultaneously and the energy storage is not required most of the time.

\subsection{Discussion}
Although the above feasibility studies are demonstrated with wind energy dispatch, VEN can exercise other applications with different kinds of energy sources and loads. In general, VEN can be considered as a medium connecting the energy sources to the loads spatially and temporally. When an energy source is not physically connected to a load (e.g., in a remote area without power infrastructure), VEN can help transport the required energy. When there is a time discrepancy between the energy supply and demand, we may advance energy from the VEN to serve the urgent loads or temporarily “store” energy in VEN to defer the delivery. However, energy transfer is a fundamental operation of VEN and thus this capability is demonstrated in Section \ref{subsec:operational}.

\begin{table}[!t]
\renewcommand{\arraystretch}{1.3}
\caption{Comparison between the power grid and VEN.}
\label{tab:compare}
\centering
\begin{tabular}{l|l|l}
\hline\hline
 & Power grid & VEN\\
\hline
\multirow{2}{*}{Functionality} 			&  \multirow{2}{*}{Traditional power delivery}		& Immature smart grid  \\
										&																& applications \\
Cost					&	 Higher	& Lower\\
Efficiency 		& Higher	& Lower\\
Practicality	& High		& High	\\
Flexibility		& Lower		& Higher\\
Marketability & Lower		& Higher\\
\hline\hline
\end{tabular}
\vspace{-0.3cm}
\end{table}
VEN and the power grid seemingly share some similarities in terms of energy/power delivery capability. However, they have many subtle differences in functionality, cost, efficiency, practicality, flexibility, and marketability. They are summarized in Table \ref{tab:compare} and further discussed below:
\subsubsection{functionality}
The power grid is highly sophisticated and its utilization is strictly regulated. It mainly serves for traditional power delivery with real-time matching between generations and loads. On the other hand, VEN is self-contained and there are less stringent regulations on the energy sources and loads. It is a good candidate to test, and serve, immature smart grid applications.
\subsubsection{Cost}
Constructing a power grid or even expanding an existing power system is extremely complicated. It involves many issues related to network topology, facility ratings, siting, right-of-way, and visual and esthetic effects \cite{windcost,SEP1,SEP2}. It may need a decade or more for the planning and the associated time cost can be considerable. On the other hand, VEN can be built by attaching to existing infrastructure, i.e., the road network. The required equipment is smaller in scale and VEN mostly takes advantages of the equipment which is designed and utilized for their original purposes (e.g., dynamic charging is primarily designed for the convenience of battery charging for basic EV operation). If the related technologies have become mature, VEN can be deployed in relatively shorter time.
\subsubsection{Efficiency}
The power grid is specially designed for power transfer. The power lines are rated for low power loss and implementing optimal power flow allows minimizing overall power loss \cite{OPF}. It imposes strict regulation to its utilization, resulting in high efficiency. On the other hand, most parts of VEN are primarily designed for other purposes but their integration can exercise energy transfer. It imposes fewer regulations allowing a broader range of applications. The frequent WPT may incur certain energy loss and its efficiency is generally lower. 
From the simulation results given in Section \ref{subsec:operational}, the system efficiency can be improved through improvements in WPT efficiency (i.e., $z$), the packet size (i.e., $w$), and EV penetration. With technological advancements, WPT efficiency will be enhanced and the packet size can be made larger. Longer (dis)charging tracks and bigger EV battery reserve can also allow larger packet sizes. EV penetration can be improved through enforcing various government green policies and promoting the public's environmental awareness. Moreover, dedicated known vehicular schedules (e.g., from public transport and autonomous vehicle systems \cite{AVPTS}) allow formation of energy paths with fewer charging and discharging events. This can reduce energy loss and thus improve the system efficiency.
\subsubsection{Practicability}
The power system was invented for more than a century. It has been kept being modified and improved and it is the most important means for power delivery. On the other hand, although VEN is primarily investigated in this paper, the development of most required equipment is promising and the corresponding technologies are likely to become mature in a few years’ time. Hence both the power grid and VEN are highly practical.
\subsubsection{Flexibility}
Following the discussion on cost above, the power grid is once constructed, it is not easy to be reconfigured. Moreover, new smart grid applications may not be easily deployed in the power system. On the other hand, the structure of VEN can be modified rather easily and it can accommodate more applications. Hence VEN has higher flexibility.
\subsubsection{Marketability}
The power grid is generally constructed and managed by a small number of parties (e.g., regional transmission organization and independent system operator).  Energy markets over the power system are generally well structured and regulated. On the other hand, various scales of VEN can be easily constructed by interested organizations. New energy markets, e.g., vehicle-to-grid energy trading \cite{V2GMarket,EVprofit}, can be deployed over VEN relatively easily. Hence, VEN achieves higher marketability.

From above, both systems have their own advantages and each has its own role of energy/power delivery. VEN can accomplish some tasks that the traditional power grid cannot do easily, and vice versa. They are not replacing one another. As a whole, VEN may be considered as a feasible design to complement the power network and enhance the overall power system performance.

\section{Conclusion} \label{sec:conclusion}

The smart grid introduces many new elements which are not easily  incorporated into the existing power system due to power reliability and security reasons. Many interesting smart grid implementations  can be more easily realized  if there exists an independent infrastructure for power delivery without interference to the existing power network. In this paper, VEN is introduced and it is capable of distributing energy across a large geographical area with EVs. VEN is built upon the traffic network and it relies on existing well-developed technologies. Small amounts of energy, as energy packets, are carried by EVs through multiple vehicular routes. By carefully designing the energy packet routing, energy can be transported from energy sources to destinations. In this paper, a preliminary study of VEN is given and a model for further analysis is developed. Its operational and economic feasibilities is studied by setting up a VEN with real road traffic data in the U.K. It is shown that a considerable amount of renewable energy can be transported from some remote wind farms to London under some reasonable assumptions. It is also revealed that VEN can be profitable in the near future. This work is not intended to posit VEN as a perfect solution for energy delivery. Instead, VEN may be considered as a feasible design to complement the power network and enhance the overall power system performance.  Besides energy delivery dedicated to a source-destination pair, it can also serve as a big energy buffer with many interfaces across a region for energy absorption and distraction. The stored energy can be utilized in a wiser way economically and operationally and this may facilitate many applications which are hard to be realized in the power grid or stationary energy storage alone. It is likely that VEN can be realized and deployed in the next decade. With the energy packet-switching design, VEN has many potential future extensions by incorporating ideas and results from the packet-switched data network.

%\section*{Acknowledgment}

%
%

\ifCLASSOPTIONcaptionsoff
  \newpage
\fi

%\bibliographystyle{IEEEtran}
%\bibliography{IEEEabrv}

%

\end{document}